\documentstyle{article}

\def\QQa{\renewcommand{\baselinestretch}{1.3}\Huge\normalsize\large\small}

\begin{document}
\QQa

\large
\begin{flushright}
ITP.SB-93-13\\

\end{flushright}

\begin{center}
\Large
{\bf  Dynamical Realizability for Quantum Measurement and Factorization
of Evolution Operator}\\
\vspace{1cm}
\Large

Chang-Pu Sun \\

Institute for Theoretical Physics,State University of New York,Stony Brook,
NY 11794-3840,USA\\
and\\
Physics Dpartment,Northeast Normal University,
Changchun 130024,P.R.China\\
\vspace{1cm}

\Large
{\bf Abstract}\\
\end{center}

\large
By building a general dynamical model  for quantum measurement process,it
is shown that the factorization of reduced evolution operator sufficiently
results in the quantum mechanical realization of the wave packet collapse
and the state correlation between the measured system and the measuring
instrument-detector.This realizability is largely independent of the details
of both the interaction and Hamiltonian of detector.  The Coleman-Hepp model
and all its generalizations are only the special cases of the more universal
model given in this letter.An explicit example of this model is finally given
in connection with coherent state.\\
---------------------------------------------------------------------------\\
{\bf PACS} number 03.65.Bz,03.80.+r

\newpage

It is well-known that,though the theory of quantum mechanics and its
applications are extremely successful,its interpretation in connection
with the corresponding measurement is still a valued problem that
the physicist must face [1-3].Since an exactly-solvable model was
presented twenty years ago [4] to describe von Neumann's wave packet collapse
(WPC) in measurement  as a quantum dynamical proccess caused by the
interaction between the measured system (S) and the measuring instrument-
detector (D),the considerable studies have been focused this model[5-9]
which is now called Coleman-Hepp (CH) model.More recently,it was
respectively generalized to the case with energy exchanging between S and D
[8]and to the case simultaneously with the classical limit -the
large quantum number limit and the macroscopic limit-the large particle number
limit[9].Notice that  another important problem in quantum measurement,the
state correlation between S and D(SCBSD)can also be  studied by making use
of other solvable-models,e.g.,the Cini model  in ref.[10].

However,all the investigations mentioned above  only proceeded with
the concrete forms of interaction and thus the main conclusions seem to
depend on the selecton of concrete form of the interactions.
It is undoubtful that the model-idenependent study for this problem
is much appreciated.The present studies
,which is briefly reported in this letter, are  dedicated to seek of essence
of the quantum machanical realization of the WPC and SCBSD
in the CH-model and its generalizations.To this end a more univrsal
dynamical model of quantum  measurement is proposed as interaction
-independently as possible.Based on this model,we will show that
the realizability of the WPC  and SCBSD as a quantum
dynamical process mainly depends on the factorizability of the
reduced evalution operator for the system.This crucial observation not
only reveals the
essence rooted in those well-establiched
 exactly-solvable models for quantum measurement,
but also provides us with a guidance to find new  exactly-solvable models.
Finally,an exactly-solvable model associated
coherent state will be  analysed as an explicit example of them in detail.
\vspace{0.4cm}

{\bf 1.The General Model}.Our model can be regarded as an universal promotion
of the original CH model.The measurd system S is sill represented by an
ultrarelativisic particle with the free Hamiltonian $H_0=c\hat p$ ,but
the detector D is made of N particles with single-particle Hamiltonian
$h_k(x_k),(k=1,2,...,N)$ which is Hermitian. S is assumed to be {\it
independetly} subjected to the interaction $V_k(x,x_k)$ of each particle k.
Here,x and $x_k$ are the coordinates of S and the single particle
k in D respectively and the k'th interaction potential $V_k(x,x_k)$
only depeds on $x$ and
$x_k$ and  $h_k(x_k)$ on the single particle coordinate $x_k$ and the
corresponding momentum.Then,we can write down the total Hamiltonian
for the `universe '= S + D
$$H=H_0+H'=H_0+H_I +H_D:$$
$$H_I=\sum^N_{k=1}V_k(x,x_k),H_D=\sum^N_{k=1}h_k(x_k),\eqno{(1)}$$
where
$$H'=H_I+H_D=\sum^N_{k=1}[h_k(x_k)+V_k(x,x_k)],\eqno{(2)}$$
is {\it a direct sum decomposation} of single-particle forms.This fact,
associated with the fact that the $H_0$ is of the first order of $\hat p$,will
lead to the factorization of effective (reduced) evalution operator and thereby
produces the WPC in quantum measurement.This factorizability is also
related to the SCBSD closely.Fortunately,
to prove it we need not the further
concrete forms of both  $h_k(x_k)$ and $V_k(x,x_k)$.In this sense we say
this model is more universal.It is worth notice
that the original HC model and its generalizations are the special
examples of this universal model.
\vspace{0.4cm}

{\bf 2.The Evolution Operator}.In order to interpretate the WPC and
SCBSD
as the consequence of the Schrodinger evolution of the universe(S+D),
we should consider the properties of the evolution operator defined
by the general Hamiltonian (1).Following Hepp [4],we first transform
into the `moving ' representation(also loosely called interacton
representation )by assuming the evolution operator to be the following
form
$$U(t)=e^{-ict\hat p/\hbar}U_e(t),\eqno{(3)}$$
Obviously,the reduced evalution operator $U_e(t)$ obeys an efficvive
Schrodinger equation
$$i\hbar \frac{\partial}{\partial t} U_e(t)=H_e(t)U_e(t),\eqno{(4)}$$
with the effective Hamiltonian
$$H_e(t)=\sum^N_{k=1}h_{ek}(t)=\sum^N_{k=1}[h_k(x_k)+V_k(x+ct,x_k)],\eqno{(5)}
$$
depending on time.
Since $H_e(t)$ is a direct sum of the time-dependent Hamiltonians  $h_{ek}(t)$
(k=1,2,...,N) parameterized by $x$,the x-depenedent evolution operator ,as
the formal solution to the eq.(4)
$$U_e(t)=\prod^N_{k=1}\otimes U^{[k]}(t)
=U^{[1]}(t)\otimes U^{[2]}(t)\otimes ....\otimes U^{[N]}(t),\eqno{(6)}$$
is factorizable,that is to say,$U_e(t)$ is a direct product of the
single-particle
evalution operator
$$U^{[k]}(t)=\Im exp[(1/i\hbar \int_{0}^{t}h_{ek}(t)dt],\eqno{(7)}$$
where $\Im$ denotes the time-order operation.As proved in as follows ,
it is just the above factorizable property of the reduced evolution
operator that results in the quantum dynamical realization of the WPC and
closely related to the SCBSD
in quantum measurement.
\vspace{0.4cm}

{\bf 3.The Wave Packet Collapse as a Quantum Dynamical process}.Let us
use the ideal double-slit experiment of interference to show the WPC
as the consequence of the quantum dynamical evolution  of
the universe S+D described by the above factorizable evolution operator.
An incident wave  is split by a divider into two branchs $\mid \psi_1>$
and $\mid \psi_2>$ and the detector D is in the ground
state
$$\mid 0>=\mid 0_1>\otimes\mid 0_2>\otimes...\mid 0_N>,\eqno{(8)}$$
in the same time where $\mid 0_k>$ is the ground state of $h_k(x_k)$.
Then,the initial
state for the universe S+D is
$$\mid \psi(0)>=(C_1\mid \psi_1>+
C_2\mid \psi_2>)\otimes \mid 0>,\eqno{(9)}$$
Notice that the ground state is required by a stable measuring instrument.
Starting with this initial state the universe S+D will evolve according
to the wave function
$$\mid \psi(t)>=C_1\mid \psi_1>\otimes \mid 0>+C_2\mid \psi_2>)\otimes
U_e(t)\mid 0>,\eqno{(10)}$$
Here,like the authors in ref.[4-8], we have supposed that only
the second branch wave $\mid \psi_2>$
interacts with D so that the double-slit experiment of interference
can be realized.In fact,such a partiality of interaction for different
states of S can be automatically
given in the `momentum'(p-) representation with the basis $\mid p>$
if we use an improved Hamiltonian
$$H=H_0+H'=H_0+H_I\delta_{p_0}(\hat p) +H_D,\eqno{(11)}$$
obtained by introducing an operator function $\delta_{p_0}(\hat p)$:
$$\delta_{p_0}(\hat p)\mid p>=\delta_{p_0,p}\mid p>$$
to the original Hamiltonian (1) and take
$$\mid \psi_1>=\sum_{p'\neq p_0} C_{p'}'\mid p'>,\mid \psi_2>=\mid p_0>,
\eqno{(12)}
$$

{}From the final state (10) we explicitly write down the density matrix for
the universe S+D

$$\rho(t)=\mid \psi(t)><\psi(t)\mid=|C_1|^2\mid \psi_1(t)><\psi_1(t)\mid
\otimes \mid 0><0\mid + $$
$$|C_2|^2|\psi_2(t)><\psi_2(t)\mid\otimes U_e(t)\mid 0><0\mid U_e(t)^+$$
$$  +C_1C_2^*\mid \psi_1(t)><\psi_2(t)\mid\otimes U_e(t)\mid 0><0\mid $$
$$+C_2C_1^*\mid \psi_2(t)><\psi_1(t)\mid)\mid\otimes \mid 0><0\mid U_e(t)^+
\eqno{(13)}$$
In the problem of WPC,
because we are only intereset in the behaviors of the system S and the
effect of the detector D on it,we only need
the reduced density matrix for S
$$\rho(t)_S=Tr_D \rho(t)=|C_1|^2\mid \psi_1(t)><\psi_1(t)\mid
+|C_2|^2\mid \psi_2(t)><\psi_2(t)\mid+$$
$$(C_1C_2^*\mid \psi_1(t)><\psi_2(t)\mid+C_2C_1^*\mid \psi_2(t)><\psi_1(t)\mid)
<0\mid U_e(t)\mid 0>,\eqno{(14)}$$
where $Tr_D$ represents the trace to the variables of the detector D.
Let us recall that the WPC postulate means the  reduction of pure state density
matrix
$$\rho(t)_S\rightarrow
\hat{\rho(t)}=|C_1|^2\mid \psi_1(t)><\psi_1(t)\mid+|C_2|^2\mid \psi_2(t)>
<\psi_2(t)\mid,\eqno{(15)}$$
Obviously,under a certain condition to be determined ,if $<0\mid U_e(t)\mid 0>
=0$,then the coherent terms in eq.(14) vanish  and the quantum dynamics
automatically leads to this reduction,i.e.,the
WPC!Now,let us prove that this condition is just the macroscopic limit
defined by very large particle number N of D,i.e.,$N\rightarrow \infty$.
In fact,due to the facterization of the reduced evolution operator
$ U_e(t)$,the norm of $<0\mid U_e(t)\mid 0>$ is
$$|<0|U_e(t)|0>|=\prod^{N}_{k=1}|<0_k\mid U^{[k]}\mid 0_k>|
= exp[-\sum^N_{k=1}\Delta_k(t)],\eqno{(16)}$$
where
$$e^{-\Delta_k(t)}=|<0_k\mid U^{[k]}\mid 0_k>|
=[1-\sum_{n\neq 0}\mid <n\mid U^{[k]}\mid 0_k>|^2]^
{1/2} \leq 1,,\eqno{(17)}$$
Usually,$\Delta_k(t)$ is a non-zero  and positive and thus the series
$\sum^{\infty}_{k=1}\Delta_k(t)$ diverges to infinity,that is to say,
$ <0|U_e(t)|0>$ as well as its norm   approach zero as $N\rightarrow \infty$.
This just proves a central conclusion
that the WPC can appear as a quantum dynamical process for the universal
model (1) in the macroscopic limit as long as the dynamical models are selected
to be have the factorizable evolution operators.However,in this case,there
is not interactions among the particles in the detector.We understand it
as an ideal case .Becuase the particles in a realistic measuring instrument
must interact with each other,it is necessary to built the exactly-solvable
dynamic model of quantum measurement with  self-interacting detector.We
belive the above -mentioned facterization property probably is also a clue
to find such model.

In terms of the above model,we can also discuss the energy-exchang process and
the delicate behaviour in $N\rightarrow \infty$ such as in ref.[8].For the
latter,we have
$$ |<0|U_e(t)|0>|\sim e^{-N\bar{\Delta}_k(t)}$$
where $\bar{\Delta}_k(t)$ represents the average value of $\Delta_k(t)$.The
above formula shows that gradual disappearence of the interference.
\vspace{0.4cm}

{\bf 4.Correlation Between States of Sytem and Detectors}.Physically,the
measurement is a scheme  using the counting number of the measuring
instrument D to manifest the state of the measured system
S.The state correlation between  S and D (SCBSD)will enjoys this manifestation.
Now ,we  show how this correlation occurs for the above dynamica model (1)
in the certain limit.To simplify the problem ,we also use the improved
Hamiltonian (11) .Let $c_{m_k}$ be a one with largest
norm among the coefficients $c_{n_k}=<n_k|U^{[k]}(t)|0_k>$(k=1,2...)
 and assume that the state $
|m_k>$ with maximum possibility amplitude is not degenerate.Then,
$$U^{[k]}(t)|0>= c_{m_k}\{|m_k>+\sum _{n\neq m}[c_n/c_{m_k}]|n>\}
.\eqno{(18)}$$
Except the coefficient of $|m_k>$,each of ones in
$c_{m_k}^{-1}U^{[k]}(t)|0>$ has a
norm less than 1 .Because of the facterization
of the reduced evolution operator,the wave function  $U(t)|0>
=\prod_{k=1}^N U^{[k]}$  will be strongly peaked  around the state
$$|m>=|m_1>\otimes |m_2>\otimes|m_3>\otimes...\otimes|m_N>.$$
If the universe S+D with the Hamiltonian (6) initially is in the state
$$|\psi(0)>=[|p_0>+|p>]\otimes |0>,p\neq p_0.$$
Then,it will evolute into a state around the  state
$$|\psi(t)>=|p_0>\otimes\prod _{k=1}^N c_{m_k} |m>  +|p>\otimes |0>.
\eqno{(19)}$$
This just menifests the correlation between the state $|p_0>$ of the system
and $|m>$ of the detector.Notice that the SCBSD can exactly
appears only for the `classical' limit,in which some paprameters or internal
quantum quantum number take their limit values
(e.g.,in ref.[9,10],this is the limit
with infinite spin).In fact,in the realistic problem,the correlation often
occurs as a good approximation valid to quite high degree.A  special
example of such  correlation problem was discussed in ref.[10].
According to the above general arguments,
We will given a new example to deal with both the WPC and SCBSD.
\vspace{0.4cm}

{\bf 4.New Exactly-Solvable Dynamical Model for Quantum Measurement and
Coherent States}.Up to now we have described the general form  of the
Hamiltonian as eq.(1) for the dynamics of quantum measurement.It is easy to see
that the CH model and its various generalizations are only some special and
explicit examples of the above more universal model
.Now,let us apply above general results ,as a guidance
rule,to built new exactly-solvable model for quantum measurement.In
this model,the WPC and SCBSD  can be simultaneously
described as a quantum dynamical process.The model Hamiltonian is
$$H=c{\hat p}+\sum^{N}_{k=1}f_k(x)[a_k^+ +a_k]+
\sum^{N}_{k=1}\hbar \omega_ka_k^+a_k,\eqno{(20)}$$
Here,the detector is made of N harmonic oscillators linearly
coupled to an  ultrarelativistic particle as the measured system;
$a_k^+$ and $a_k$
are the creation and annihilation operators for boson states respectively.The
coupling function  $f_k(x)=c_kx$ only  depends on the coordinate of
the ultrarelativistic paticle.Because the effective Hamiltonian
$$H_e(t)=\sum^{N}_{k=1}(f_k(x+ct)[a_k^+ +a_k]+\hbar \omega_k
a_k^+a_k),\eqno{(21)}$$
is completely decomposible,the reduced evolution operator $U_e$ is
factorizable  ,i.e.,$U_e=\prod_k U^{[k]}(t)$,and its factors are[11]
$$ U^{[k]}(t)=e^{-i\omega_k a_k^+a_k}e^{h_k(t)}e^{A_k(t)a_k^+ }
e^{B_k(t)a_k},\eqno{(22)}$$
where the functions $h_k(t),A_k(t),B_k(t)$ are defined by
$$A_k(t)=B_k(t)^*=-\frac{c_k}{\hbar \omega_k^2 }[(x+ct+ic/\omega_k)
e^{i\omega_k t}-ic/\omega_k-x],\eqno{(23)}$$
$$h_k(t)=\int^t_0 A_k(s)\frac{\partial B_k(s)}{\partial s}ds$$

Notice that the real part of $h_k(t)$ is $-\eta_k(t)$:
$$\eta_k(t)=\frac{c_k}{\hbar\omega_k^2}
 [(c^2t^2+xct+2(x^2+c^2/\omega_k^2)sin^2\omega_k/2-c\omega_k t(x cos\omega_kt
-(c/\omega_k)sin\omega_kt].\eqno{(24)}$$
Since $\eta_k(t)$ is larger than zero after an interval of time,the norm
$$|<0|U(t)|0>|=e^{-\sum_{k=1}^{N}\eta_k(t) }$$
 must approach zero as $N\rightarrow \infty$.This implies the dynamical
realization of the WPC for quantum measurement.

Let us show how the correlation between the states of the system
and the detector appears as a quantum dynamical process.If
the detector is initially
in its ground state $\mid 0>=\mid 0_1>\otimes\mid 0_2>\otimes...\mid 0_N>$,the
state at t is a direct product of the coherent states.

$$|\psi^k(t)>= U^{[k]}(t)|0>=e^{h_k(t)}\sum_{n=0}^{\infty}
e^{-in\omega_kt}\frac{A_k^n(t)}
{n!}a_k^{+n}|0_k>,\eqno{(25)}$$
Using the Stirling formula,we immediately determine the value $\bar n_k$ of
 quantum number n
for which the norm of the coefficient of Fock state $|n>_f=1/(n!)^{1/2}
a_k^{+n}|0_k>$ in  the above expansion (25) is maximum,obtaining
$$\bar{n_k}=|A_k(t)|^2, \eqno{(26)}$$
In this time the validity of Stirling formula require
$\bar{n_k}=|A_k(t)|^2$ to be sufficiently large.
This means that the counting number
of the detector is macroscopically large. It is just what we expect for a
measuring instrument.If we take the initial state of the system is
$|\psi(0)>=[v|p_0>+w|p\neq p_0>]\otimes |0>$,then the correlation is
enjoyed by the Schrodinger evoluting state
$$|\psi(t)>=v|p_0>\otimes |\bar{n_1}>\otimes|\bar{n_2}>\otimes ...
\otimes |\bar{n_N}>+w|p>\otimes |0>.\eqno{(27)}$$
\vspace{0.4cm}

{\bf 6 Final Remarks}.Finally,we should point out that in practical problems
there must exist interactions among the particles constituting the detector
D.They seem to break the factorization of the reduced evolution operator.How
to realize the quantum measurement both for the WPC and SCBSD in this case
is an open question we must face.It is expected ,at least for some
special case ,that the certain canonical (or unitary)
transformation possibly enable
these particles to become the quasi-free ones .This is just similar to
 the the system
of harmonic osscilators with quadric coupling.In this case ,we can imagine that
the detector is made of free quasi-particles that do not interact with each
other.If each quasi-particle interacts with the system independently,then
the factorizability of the evolution operator can perseveres in the solvable
models for quantum measurement.

We also remarks on the realization of the double-slit type expriment where
the interaction selects only one of the two branch wave functions.For the
introduction of $\delta_{p_0}(\hat p)$ in the Hamiltonian (11)(to realize
the partiality of interaction),someone may
not fell content.In fact,we can also enjoy this selection in a quite natural
way.If the system is a spin-1/2 with the free Hamiltonian $H_0=\hbar \omega
\sigma_3$ and the detector is still defined by the general Hamiltonian
in eq.(2),the following interaction
$$H_I=g(1+\sigma_3)\sum^{N}_{k=1}h_k(x_i),\eqno{(28)}$$
 naturally results in the selection of interaction.Namely,the detector
only acts on the spin-up state .The spin-down state is free of interaction.
Thus,the eq.(28)
defines new  dynamical model for quantum measurement ,which is an
extensive generalization of Cini's model [10].
\vspace{0.5cm}

The author wishs to express his sincere thanks to Prof.C.N.Yang for drawing his
attentions to new progress in the quantum measurement .He also thank D.H.
Feng,R.M.Xu ,L.H.Yu and W.M.Zhang for many useful discussions.
He is supported by Cha Chi Ming fellowship through the CEEC in State
University of New York at Stony Brook and in part by the
NFS of China through Northeast Normal University .

\newpage

{\bf References}
\begin{enumerate}
\item J.von Neumann,{\it Mathematische Grundlage de Quantummechanik,}
Springer, Berlin,1933.
\item V.B.Braginsky,F.Y.Khalili,{\it Quantum Measurement},Cambridge,1992.
\item R.Omnes,Rev.Mod.Phys.64(1992),339.
\item K.Hepp,Helv.Phys.Acta.45(1972),237.
\item J.S.Bell,Helv.Phys.Acta.48(1975),93.
\item T.Kobayashi,Found.Phys.Lett.5(1992),265.
\item M.Namik,S.Pascazio,Phys.Rev.A44.(1991),39.
\item H.Nakazato,S.Pascazio,Phys.Rev.Lett.70(1993) 1.
\item C.P.Sun,{\it Influence of detector on quantum interference
in macroscopic and classical limits,} NENU preprint,1992,to
be published in Commu.Theor.
Phys.
\item M.Cini,Nuovo Cimento,73B(1983),27.
\item W.M.Zhang,D.H.Feng,R.Gilmore,Rev.Mod.Phys.62(1990) ,867.

\end{enumerate}
\end{document}